# Charge carrier coherence and Hall effect in organic semiconductors.

01/13/2016

H. T. Yi[1], Y. Gartstein[2] and V. Podzorov[1,3]

[1] Dept. of Physics, Rutgers University, Piscataway, NJ 08854, USA;  [2] Dept. of Physics, University of Texas at Dallas, Richardson, TX, USA;  [3] Inst. for Adv. Mater. & Devices for Nanotech., Rutgers University, Piscataway, NJ 08854, USA.

[*] Corresponding author:  podzorov@physics.rutgers.edu

Hall effect measurements are important for elucidating the fundamental charge transport mechanisms and intrinsic mobility in organic semiconductors.  However, Hall effect studies frequently reveal an unconventional behavior that cannot be readily explained with the simple band-semiconductor Hall effect model.  Here, we develop an analytical model of Hall effect in organic field-effect transistors in a regime of coexisting band and hopping carriers.  The model, which is supported by the experiments, is based on a partial Hall voltage compensation effect, occurring because hopping carriers respond to the transverse Hall electric field and drift in the direction opposite to the Lorentz force acting on band carriers.  We show that this can lead in particular to an underdeveloped Hall effect observed in organic semiconductors with substantial off-diagonal thermal disorder.  Our model explains the main features of Hall effect in a variety of organic semiconductors and provides an analytical description of Hall mobility, carrier density and carrier coherence factor.

Unique combination of processability, interesting optoelectronic properties and mechanical flexibility makes molecular semiconductors very promising for development of organic electronics. However many fundamental aspects of charge carrier transport in these materials are still not fully understood. Weak van der Waals intermolecular interactions in these materials may lead to formation of rather narrow (0.1 - 0.3 eV) electronic bands of extended states[1], that can be relatively easily destroyed by thermal molecular fluctuations[2,3]. Such fluctuations in molecular crystals can create significant off-diagonal disorder, leading to localized states in the gap below the mobility edge (the tail states)[2-4]. Thus, the intrinsic factors that compete in determination of the dominant transport mechanism in organic semiconductors include intermolecular interactions, governed by the equilibrium molecular positions in the crystal (the transfer integrals)[1], carrier self-localization due to polaron formation (local electron-phonon coupling)[5,6], and off-diagonal thermal disorder (non-local electron-phonon coupling)[2,3]. This leads to a host of small-molecule organic semiconductors with the intrinsic (that is, not dominated by static disorder) charge transport mechanism varying from an incoherent hopping in localized states to a pure band transport in extended states (see, e.g., [7] and refs. therein). For instance, several high-performance organic semiconductors, with a band transport and carrier mobilities in the range $\mu \sim 1 - 20$ cm$^2$V$^{-1}$s$^{-1}$, have been recently identified[7]. On the other hand, there are crystals that seem to be at the borderline between a coherent band transport and an incoherent hopping. Thermal disorder, undoubtedly detrimental for robust band transport, is affected by specific molecular structure and crystal packing, and it can be suppressed at low temperatures[2-4]. Besides these intrinsic factors, static disorder (chemical impurities and structural defects) also plays an important role in organic semiconductors by leading to in-gap trap states that immobilize charge carriers at various time scales[8,9].

It is important to emphasize that localized tail states in pristine crystalline organic semiconductors can originate just from off-diagonal thermal disorder (thermal fluctuation of molecules with respect to each other). Such tail states are therefore to be considered as an intrinsic phenomenon occurring even in defect-free highly ordered single crystals. The important distinction, however, between thermally-induced and defect-induced disorder is that the former has a dynamic and homogeneous character (that is, it occurs uniformly throughout the entire crystal at every lattice site), while the latter is "frozen" at specific diluted spatial locations. The concentration of physical defects in high-purity molecular crystals is usually much smaller than the density of molecules. For instance, molecular density of crystalline rubrene is $1.43 \times 10^{21}$ cm$^{-3}$, while the density of rubrene peroxide defects at the surface of severely photooxidized crystals is $\sim 0.2 \times 10^{21}$ cm$^{-3}$ and only $\sim 0.7 \times 10^{19}$ cm$^{-3}$ in pristine crystals (in the bulk, this concentration is even lower)[10]. Thus, the density of tail states due to thermal disorder in such crystals is expected to be much greater than that due to physical defects, and therefore thermal disorder should play a dominant role in determining the charge transport mechanism in these materials (provided, of course, that it wins over the π-π intermolecular interactions). Based on this consideration, one would anticipate that materials, whose charge transport is dominated by thermal disorder (or, conversely, by band forming π-π interactions), will likely continue to behave as hopping (or, band-like) systems, even if an additional static disorder is introduced. Such disorder can be intentionally added in the form of chemical impurities or structural defects, produced, for instance, by photooxidation[10,11], absorption of atmospheric gasses[12,13], addition of molecular impurities during the crystal growth[14], or ionizing radiation[15]. On the contrary, variation of temperature should have a much stronger effect on the transport properties of materials dominated by off-diagonal thermal disorder.

A conventional band-semiconductor Hall effect is an important signature in the studies of intrinsic charge transport in organic semiconductors, because its observation signals the presence of delocalized band-like states. Indeed, the classic Lorenz force considerations in the Hall bar geometry are readily applicable to delocalized band carriers (that is, carriers that have a well-defined microscopic drift velocity), but not to substantially localized carriers[7,16]. In such a simplistic picture, Hall effect would mainly probe instantaneously mobile charge carriers in extended states, thus yielding their concentration, $n_{Hall}$, and intrinsic trap-free mobility, $\mu_{Hall}$. It is then remarkable that Hall effect measurements in organic field-effect transistors (OFETs) frequently lead to a surprising observation of not well-understood *improper* Hall effect, in which Hall carrier density, $n_{Hall} \equiv (e \cdot R_H)^{-1}$, where e is the electron charge and $R_H$ is a Hall coefficient (see below), appears to be greater than the total carrier density electrostatically induced in OFETs (excluding those trapped in deep traps), $n_{FET} \equiv e^{-1} \cdot C_i(V_G - V_{th})$, where $C_i$ is a gate-channel capacitance per unit area, $V_G$ is the gate voltage, and $V_{th}$ is a threshold voltage in the linear-regime measurements of OFETs[17]. Simultaneously, Hall mobility, $\mu_{Hall}$, appears to be smaller than the longitudinal drift mobility, $\mu_{FET}$ ($n_{Hall} > n_{FET}$, $\mu_{Hall} < \mu_{FET}$). This behavior, which is also sometimes called an *underdeveloped* Hall effect, contradicts to the simple interpretation of Hall effect as probing only instantaneously mobile charge carriers and therefore giving the highest possible, trap-free mobility ($\mu_{Hall} \geq \mu_{FET}$) [7]. An improper Hall effect has recently been observed in several systems, including pentacene[17], 6,13-bistriisopropyl-silylethynyl pentacene (TIPS-pentacene)[18], and even conjugated polymers[19,20]. It was empirically assigned to a partial carrier coherence, caused in these systems by strong off-diagonal thermal disorder, and parameterized by a carrier coherence factor, $\alpha$, defined as the ratio of the total carrier density to that determined from Hall effect measurements, $\alpha \equiv n_{FET}/n_{Hall}$ [17].

The postulated empirical relation between the carrier coherence factor and improper Hall effect does not offer a mechanistic picture of the phenomenon. It should also be noted that recently reported high-resolution *ac* Hall measurements have led to an observation which cannot be easily rationalized within that empirical framework. The *ac* technique allows for reliable measurements of Hall mobility and carrier density in OFETs with very low drift mobilities $\mu < 1$ cm$^2$V$^{-1}$s$^{-1}$, in which conventional Hall effect measurements are hardly possible even in very high *dc* magnetic fields of up to 12 T[21]. The *ac* Hall effect studies show that certain crystals expected to have a strong thermal disorder at room temperature, such as for instance tetracene[22], still exhibit a normal (fully developed) Hall effect, even though their carrier mobility is as low as $\mu \sim 0.3$ cm$^2$V$^{-1}$s$^{-1}$ [21]. This observation thus does not support the view that less coherent carriers *always* lead to an underdeveloped Hall effect with $\alpha < 1$.

Here, we propose a simple picture and the corresponding model description that could readily accommodate many salient features observed in recent Hall effect studies in organic semiconductors. The picture is based on the coexistence of band-like (delocalized) and hopping (localized) charge carriers in the accumulation channel of OFETs that respond differently to the applied magnetic field. Specifically, while the delocalized carriers exhibit the conventional Lorentz-force effect, the localized carriers are assumed to have a negligible response. It should be mentioned that Hall effect in purely hopping transport regime, originating from quantum-mechanical interference effects on the closed-loop trajectories, was theoretically predicted[23], but apparently thus far has eluded clear experimental observations[24]. One may speculate that local hopping trajectories in systems with strong off-diagonal disorder can have a more one-dimensional character, thereby substantially decreasing the loop interference effects in comparison with those predicted for hopping carriers on regular lattices. This assumption appears to be an interesting subject for a dedicated theoretical study, which is beyond the scope of this paper. Of course, the hopping carriers would still experience a transverse Hall electric field, produced by the band carriers, and drift in the transverse direction opposite to the Lorentz force acting on the band carriers (Fig. 1). In equilibrium, the resulting Hall voltage is correspondingly modified by the hopping carriers. The resultant equations derived below, relating Hall effect and FET mobilities and carrier densities, appear to capture most of the recently observed experimental trends.

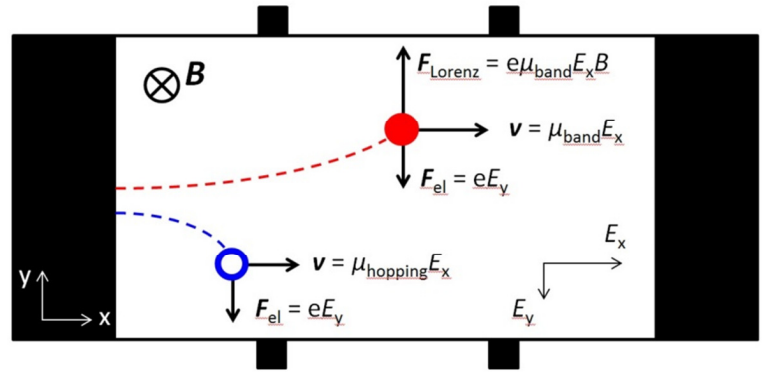

**Figure 1.** The two types of charge carriers (both holes) in organic semiconductors with disorder: a coherent band-like hole moving in delocalized states (solid red circle) and a hopping hole moving in localized tail states (open blue circle). In Hall effect measurements, these carriers are skewed towards the opposite sides of the channel, because the band carriers experience both the Lorentz and electric forces, while the hopping carriers only respond to the electric forces. The motion of hopping carriers in the transverse direction leads to a (partial) compensation of the Hall voltage, which results in underestimated mobility and overestimated carrier density in Hall effect measurements, as compared to those obtained in longitudinal FET measurements ($\mu_{Hall} < \mu_{FET}$, $n_{Hall} > n_{FET}$).

For clarity, let us start with a simple textbook discussion of an accumulation channel comprising only mobile band carriers with density $n$ and mobility $\mu$. In the geometry of Hall effect measurements in OFETs, these carriers experience a Lorentz force, $|F_L| = |e[v \times B]| = evB$, where $v = \mu \cdot E$ is a longitudinal terminal drift velocity of the carrier motion, occurring with a drift mobility $\mu$ in the longitudinal source-drain electric field $E = V/L$, $V$ is the source-drain voltage applied to the accumulation channel of length $L$, and $B$ is the magnetic field perpendicular to the channel. This force causes the carrier trajectories to bend toward one side of the channel, leading to a Hall voltage, $V_{Hall}$, developing across the channel. The transverse Hall electric field, $V_{Hall}/W$ (W is the channel width), balances the Lorentz force, thus leading to an equilibrium according to the equation: $e\mu(V/L)B = eV_{Hall}/W$. Here, $V/L$ and $V_{Hall}/W$ are the two orthogonal electric fields corresponding to the source-drain and Hall voltages, respectively. Such a system is described by the following relationship between Hall voltage $V_{Hall}$, source-drain (longitudinal) current $I$, magnetic field $B$, and mobile carrier density $n$:

$$V_{Hall} = \frac{I \cdot B}{en} \equiv R_H \cdot I \cdot B = \left(\frac{W}{L}\right)\left(\frac{1}{en}\right)\sigma VB, \qquad (1)$$

where $\sigma \equiv \sigma_x = en\mu$ is the longitudinal channel conductivity per square. It can be rewritten as:

$$V_{Hall} = \frac{I \cdot B}{en} = R_H \cdot I \cdot B = \left(\frac{W}{L}\right)\mu VB. \qquad (2)$$

These formulas are normally used to extract the Hall carrier mobility, $\mu_{Hall}$, and Hall carrier density, $n_{Hall}$, from experiments as:

$$\mu_{Hall} = \left(\frac{L}{W}\right)\left(\frac{V_{Hall}}{V}\right)\frac{1}{B} = \frac{E_y}{E_x B} \quad \text{and} \quad en_{Hall} = \left(\frac{W}{L}\right)\left(\frac{V}{V_{Hall}}\right)\sigma B = \left(\frac{E_x}{E_y}\right)\sigma_x B = \frac{\sigma_x}{\mu_{Hall}} \qquad (3)$$

Note that in these equations, the current $I$, conductivity $\sigma$ (or $\sigma_x$) and carrier density $n$ are meant to represent *only* the mobile band carriers, because only these carriers experience the Lorentz force. If the transport is purely band-like, $\mu_{Hall}$ and $n_{Hall}$ determined from Hall measurements according to Eq. 3 would then represent the actual mobility and density of mobile carriers. In a mixed transport regime of our model, however, when both band and hopping carriers are present in the accumulation channel and contribute to the longitudinal conductivity, direct application of Eq. 3 might result in an error. Indeed, the longitudinal conductivity in this case is:

$$\sigma = \sigma_{band} + \sigma_{hopping} = en_{band}\mu_{band} + en_{hopping}\mu_{hopping}, \qquad (4)$$

where $\sigma_{band}$ and $\sigma_{hopping}$ are the portions of conductivity due to band and hopping carriers, with the corresponding densities and mobilities of these carriers, $n_{band}$, $n_{hopping}$ and $\mu_{band}$, $\mu_{hopping}$. This shows that using the total conductivity $\sigma$ or the total current $I$ in Eqs. 1-3 might lead to at least one problem: an overestimated mobile carrier density. For illustration of such an error, consider applying Eq. 1 to a case of pure hopping conduction, when $V_{Hall} = 0$ and $I \neq 0$, which leads to an erroneous result $n \rightarrow \infty$. This would not occur, however, if one used the correct value for $I$, the current carried by mobile band charges, which is zero in this example. There are further important considerations to be taken into account as we describe below.

Despite the absence of a transverse magnetic field effect on hopping carriers (within the model discussed in this work), these carriers should still affect the total Hall voltage presented across the channel, because they do "feel" the transverse electric field, $E_y \equiv V_{Hall}/W$, associated with the Hall voltage. This field pushes the hopping carriers in the direction opposite to the Lorentz force and results in the corresponding compensation of the electric charge at the opposite side of the channel (Fig. 1). This will cause Hall voltage to be smaller than it would be in the case of only band carriers present in the channel. Such a compensation effect can be significant, if sufficient number of hopping carriers coexist with band carriers (see below). It can be easily seen from Eq. 3 that this compensation of $V_{Hall}$ would lead to an underestimated mobility, that is, $\mu_{Hall}$ will be smaller than the actual mobility of band carriers, $\mu_{Hall} < \mu_{band}$. It can also be seen from Eq. 3 that in such a system, the Hall carrier density will be overestimated and in principle can be greater than the total carrier density (see below). Indeed, in the case of a mixed transport regime, one cannot experimentally identify what fraction of longitudinal channel conductivity $\sigma$ is due to band carriers and what is due to hopping carriers, and thus only the total longitudinal conductivity $\sigma$ can be used for parameter extraction from Hall effect measurements. In addition, we have no choice but to use the experimental $V_{Hall}$, which is already reduced due to the compensation effect described above. Thus, Eqs. 1-3 yield:

$$n_{Hall} = \left(\frac{W}{L}\right)\left(\frac{V}{V_{Hall}}\right)B(n_{band}\mu_{band} + n_{hopping}\mu_{hopping}) =$$

$$= n_{band}\left(\frac{\mu_{band}}{\mu_{Hall}}\right) + n_{hopping}\left(\frac{\mu_{hopping}}{\mu_{Hall}}\right) \quad (5)$$

In Eq. 5, as we showed above, the first ratio is greater than unity, $\mu_{band}/\mu_{Hall} > 1$, while the second ratio is usually much smaller than unity, $\mu_{hopping}/\mu_{Hall} \ll 1$. Thus, depending on the particular combination of the relative densities and mobilities of band and hopping carriers, $n_{band}/n_{hopping}$, and $\mu_{band}/\mu_{hopping}$, the resultant carrier density determined from Hall measurements within conventional model (Eqs. 1-5) might be greater than the actual total carrier density. We remind that in OFETs, the total gate-induced carrier density in the accumulation channel, excluding those trapped in deep trap states, is: $n_{total} \equiv e^{-1}C_i(V_G - V_{th})$ [7]. This compensatory effect might lead to an improper (underdeveloped) Hall effect.

For a self-consistent analysis of the problem, we will now suppose that accumulation channel contains two types of charge carriers that have different carrier densities and mobilities: $n_1$ and $\mu_1$ for the 1st type, and $n_2$ and $\mu_2$ for the 2nd type, respectively. Let's also assume for certainty that carriers of the 2nd type have a smaller mobility, $\mu_2 < \mu_1$. If both types of carriers experienced the conventional Lorentz and electric forces in the geometry of Hall effect measurements, the Lorentz force acting on the 2nd type of carriers would be smaller, and these carriers would contribute to the transverse current in the direction *opposite* to the transverse current flow of the 1st type of carriers. The transverse current densities corresponding to these two types of carriers can be written as: $j_{1y} = -\sigma_1 \cdot E_y + \sigma_1 \cdot F_L/e = en_1 \cdot \mu_1 \cdot (-E_y + \mu_1 E_x B)$ and $j_{2y} = -\sigma_2 \cdot E_y + \sigma_2 \cdot F_L/e = en_2 \cdot \mu_2 \cdot (-E_y + \mu_2 E_x B)$, where $E_x = V/L$ and $E_y = V_{Hall}/W$ are the longitudinal source-drain and the transverse Hall electric fields [25]. The dynamic equilibrium in the channel is achieved, when the total transverse current density is zero: $j_{1y} + j_{2y} = 0$.

This would give a relationship between the Hall electric field $E_y$ and the longitudinal current density $j_x$ for the case of coexisting two types of carriers:

$$\frac{E_y}{j_x} = \frac{B}{e} \cdot \frac{n_1 \mu_1^2 + n_2 \mu_2^2}{(n_1 \mu_1 + n_2 \mu_2)^2} \qquad (6)$$

In the limiting case of a single type of carriers with an equivalent total carrier density $n \equiv n_1 + n_2$ and band mobility $\mu_1$, this formula gives the familiar result for the Hall effect in a band-like conductor: $E_y = \mu_1 B E_x$.

In the picture we explore in this work, however, the 2nd type of charge carriers does not respond to the magnetic field. In such a case, balancing the transverse currents $j_{1y}$ and $j_{2y}$ leads to a modified formula (6):

$$\frac{E_y}{j_x} = \frac{B}{e} \cdot \frac{n_1 \mu_1^2}{(n_1 \mu_1 + n_2 \mu_2)^2}, \qquad (6a)$$

which can be rewritten as:

$$E_y = \left( \frac{1}{1 + \frac{n_2 \mu_2}{n_1 \mu_1}} \right) \cdot \mu_1 B E_x. \qquad (7)$$

We note that the Hall electric field $E_y$ given by Eq.(7) is *always* smaller than $E_y = \mu_1 B E_x$ in a purely band transport regime with the same total carrier concentration $n \equiv n_1 + n_2$, irrespectively of particular values of carrier densities and mobilities of the two components (as long as $\mu_2 < \mu_1$). The measured Hall mobility in a system of mixed band and hopping carriers will then become:

$$\mu_{Hall} \equiv \frac{E_y}{E_x B} = \mu_1 \cdot \frac{1}{1 + \frac{n_2 \mu_2}{n_1 \mu_1}} = \mu_1 \cdot \frac{1}{1 + \beta\left(\frac{1}{\gamma} - 1\right)} = \mu_1 \cdot \frac{\gamma}{\gamma - \gamma\beta + \beta}, \qquad (8)$$

which is evidently *always* smaller than the intrinsic mobility of the band-like carriers, $\mu_1$. Here we introduced two dimensionless parameters: $\gamma \equiv n_1/(n_1 + n_2) = n_1/n$ ($0 < \gamma < 1$) that represents the fraction of band-like carriers, and $\beta \equiv \mu_2/\mu_1$ ($0 < \beta < 1$) that gives the ratio of hopping and band mobilities. The corresponding carrier density extracted from Hall measurements is:

$$n_{Hall} \equiv \frac{E_x B}{E_y} \cdot \frac{\sigma_x}{e} = \frac{n_1 \mu_1 + n_2 \mu_2}{\mu_{Hall}} = n_1 \cdot \left(1 + \frac{n_2 \mu_2}{n_1 \mu_1}\right)^2 = n \cdot \gamma \left[1 + \beta\left(\frac{1}{\gamma} - 1\right)\right]^2 = n \cdot \frac{(\gamma - \gamma\beta + \beta)^2}{\gamma} \qquad (9)$$

To compare the results of Hall and FET measurements, consider now the longitudinal channel conductivity, $\sigma = e n_1 \cdot \mu_1 + e n_2 \cdot \mu_2 = e n \cdot \mu_1 \cdot (\gamma - \gamma\beta + \beta)$, where the total carrier density $n$ induced by application of a gate voltage, $V_G$, is given by $en \equiv en_{FET} = C_i(V_G - V_{th})$. The fraction of band carriers $\gamma \equiv n_1/n$ can be considered roughly independent of (or weakly dependent on) $V_G$. Of course, in real systems with high density of localized tail states, this may not be fulfilled precisely, as the relative fraction of mobile carriers may increase with gate voltage. However, in the cases when super-linearities in FET's transfer characteristics are not too strong, we can assume that $\gamma$ is almost $V_G$ independent. In such a case, longitudinal FET mobility $\mu_{FET}$ extracted from transistor measurements can be expressed through $\mu_1$, $\gamma$ and $\beta$ as:

$$\mu_{FET} \equiv d\sigma/d(en) = \mu_1 \cdot (\gamma - \gamma\beta + \beta). \quad (10)$$

It is easy to see from Eqs. (8-10) that:

$$\frac{n_{Hall}}{n_{FET}} = \frac{\mu_{FET}}{\mu_{Hall}} = \frac{(\gamma - \gamma\beta + \beta)^2}{\gamma}, \quad (11)$$

which shows that there is an inverse relationship between the ratio of Hall and FET carrier densities and that of the corresponding mobilities: an overestimation of the carrier density is linked to an underestimation of mobility, or vice versa.

Next, we show that depending on the combination of parameters, $\mu_1$, $\mu_2$, $n_1$ and $n_2$ (or, $\gamma$ and $\beta$) in Eqs. 8-10, the Hall carrier density, $n_{Hall}$, can be greater, comparable to, or smaller than the total carrier density, $n$, while the corresponding Hall mobility, $\mu_{Hall}$, can be smaller, comparable to, or greater than the FET mobility, $\mu_{FET}$, respectively. Indeed, these three cases can be derived directly from the above equations:

(a) "Strong hopping contribution", with a high concentration of hopping carriers $n_2$ and relatively high hopping mobility $\mu_2$ (that is, small $\gamma$, and/or not too small $\beta$):

$$\text{at } \frac{1}{\beta} < 1 + \frac{1}{\sqrt{\gamma}} \text{ (or, } \frac{\mu_1}{\mu_2} < 1 + \sqrt{\frac{n}{n_1}}\text{), we have } n_{Hall} > n_{FET} \text{ and } \mu_{Hall} < \mu_{FET} < \mu_1 \quad (12a)$$

(b) "Intermediate hopping contribution" (excluding the trivial solution, when all carriers are band carriers):

$$\text{at } \frac{1}{\beta} = 1 + \frac{1}{\sqrt{\gamma}} \text{ (or, } \frac{\mu_1}{\mu_2} = 1 + \sqrt{\frac{n}{n_1}}\text{), we have } n_{Hall} = n_{FET} \text{ and } \mu_{Hall} = \mu_{FET} < \mu_1 \quad (12b)$$

(c) "Weak hopping contribution", with a relatively low concentration $n_2$ and small mobility $\mu_2$ of hopping carriers (that is, not too small $\gamma$, and/or small $\beta$):

$$\text{at } \frac{1}{\beta} > 1 + \frac{1}{\sqrt{\gamma}} \text{ (or, } \frac{\mu_1}{\mu_2} > 1 + \sqrt{\frac{n}{n_1}}\text{), we have } n_{Hall} < n_{FET} \text{ and } \mu_{FET} < \mu_{Hall} < \mu_1 \quad (12c)$$

These conditions directly follow from Eqs. 8-10.

For an example of the case (12a), if the band-like carriers constitute only 1% of the total carrier population ($\gamma = 0.01$), and the mobility of the hopping carriers is 0.13 of the band mobility ($\beta = 0.13$), the measured Hall carrier density according to Eq. 9 will be about twice the total carrier density, $n_{Hall} \approx 2n$, the longitudinal FET mobility according to Eq. 10 will be $\mu_{FET} \approx 0.14\mu_1$, while the Hall mobility according to Eq. 8 will be $\mu_{Hall} \approx 0.071\mu_1$, that is, $\mu_{Hall}$ will be almost twice smaller than the longitudinal FET mobility, $\mu_{Hall} \approx \frac{1}{2} \cdot \mu_{FET}$.

For an example of the case (12b), if the band carriers are 50 times more mobile than the hopping ones ($\beta = 0.02$, which can occur for instance when $\mu_1 = 5$ and $\mu_2 = 0.1$ cm$^2$V$^{-1}$s$^{-1}$), but their number is small, say $\gamma = n_1/n = 4.16 \times 10^{-4}$, then the above equations yield $\mu_{FET} \approx \mu_{Hall} \approx 0.2$ cm$^2$V$^{-1}$s$^{-1}$ < $\mu_1$, and $n_{Hall} \approx n_{FET}$. That is, from the Hall measurements it will look like there is a good match between FET and Hall mobilities and carrier densities, but in fact the measured mobilities are far lower than the intrinsic band mobility.

Finally, for an example of the case (12c), if $\gamma = 0.25$ (a quarter of carriers are band-like) and $\beta = 0.1$ (the band-like carriers are 10 times more mobile than hopping ones), we would have $\mu_{Hall} = 0.77\mu_1$, $\mu_{FET} = 0.325\mu_1$, and $n_{Hall} = 0.42 n_{FET}$.

Note that case (12b), $n_{Hall} \approx n_{FET}$ and $\mu_{Hall} \approx \mu_{FET}$, can be easily confused with the truly *intrinsic band transport*, in which all the carriers are band-like, and there is a good correspondence between Hall effect and FET measurements (such as, for instance, in pristine rubrene OFETs [16,26]). It is clear that case (12b) is far from the intrinsic fully band-like regime, and it is still considerably affected by hopping ($\mu_{Hall} \approx \mu_{FET} < \mu_1$).

The above consideration shows that Hall effect measurements in a mixed transport regime may result in an incorrect estimate of the charge carrier mobility and density. Hall mobility can be underestimated, compared to the true mobility of band-like carriers (as well as relatively to the longitudinal FET mobility), while the Hall carrier density can be overestimated. In particular, the Hall carrier density can appear to be even greater than the total carrier density. This situation, sometimes reported in the literature, corresponds to the case of a "strong hopping contribution" described by Eq. 12a. It has to be noted that in samples with substantial intrinsic or extrinsic disorder, we are likely to have a continuous mobility distribution, instead of just two

"discrete" types of carriers, fast and slow, as we considered above. Thus, the simplified formulas derived here may not necessarily describe the actual disordered systems precisely. However, this model correctly captures the tendencies observed in Hall effect measurements in such systems. This simple result emphasizes the caution that should be exercised while interpreting Hall effect data in disordered systems with coexisting band and hopping carriers.

It would be instructive to plot the function $f(\gamma, \beta) = n_{Hall}/n_{FET} = \mu_{FET}/\mu_{Hall}$ given by Eq. 11 as a 3D plot of two parameters, $\gamma$ and $\beta$ (the Origin file with this function can be downloaded from Supplementary Materials). The function can be rotated around three axes by clicking on the plot and dragging it with a mouse, which gives a good idea about the 3D shape of the function. Fig. 2a shows the function (the blue surface) and the constant $n_{Hall}/n_{FET} = \mu_{FET}/\mu_{Hall} = 1$ (the pink horizontal plane) given for comparison. The curve, along which the blue $f(\gamma, \beta)$ surface and the pink $f = 1$ plane intercept, describes the conditions in Eq. 12b, for which Hall effect gives exactly the same result as FET measurements ($n_{Hall}/n_{FET} = \mu_{FET}/\mu_{Hall} = 1$).

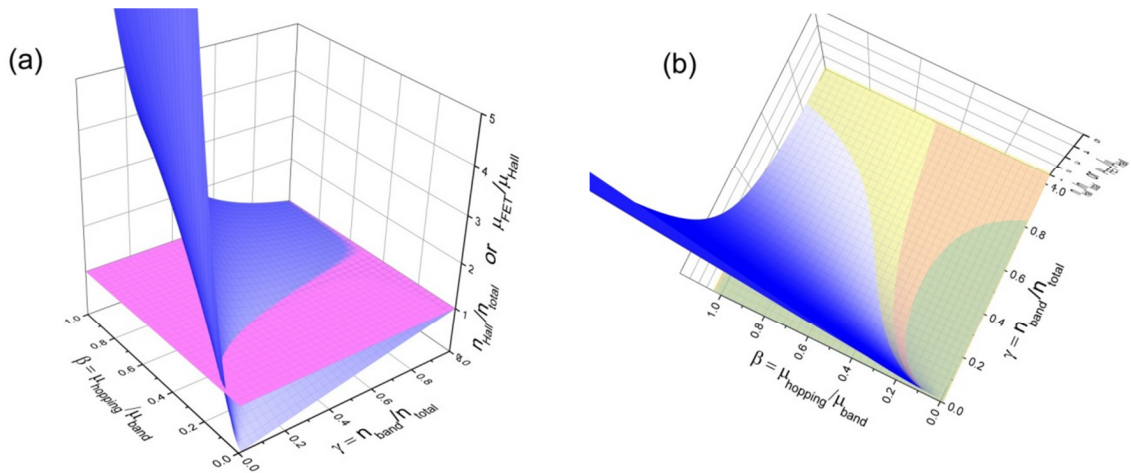

**Fig. 2. The function $f(\gamma, \beta) = n_{Hall}/n_{FET} = \mu_{FET}/\mu_{Hall}$ given by Eq. 11.** The two variables $\gamma = n_1/n$ and $\beta = \mu_2/\mu_1$ represent the fraction of band-like carriers and the ratio of hopping and band mobilities, respectively. (a) A side view of this 3D function (blue) with an added horizontal plane $f(\gamma, \beta) = 1$ (pink). The intercept of the blue and pink surfaces corresponds to the conditions given in Eq. 12b that describes a perfect match between Hall and longitudinal FET measurements. (b) A top view of the same function, with two additional planes $f(\gamma, \beta) = 1 \pm 0.2$. The two new intercepts define the parameter space, where the discrepancy between Hall effect and longitudinal FET measurements is smaller than 20% (yellow and orange areas). Origin file with function $f(\gamma, \beta)$ can be downloaded from Supplementary Materials.

We note that due to the specific shape of this function, a rather good correspondence between Hall effect and FET measurements is observed not only for the parameters along this intercept, but in a much broader parameter space. Indeed, the blue surface grazes very closely along the horizontal pink plane at large $\gamma = n_1/n$, while going almost perpendicular to it for small $\gamma$ and $\beta$ near the origin. This shows that there is a large parameter space, where the function $f(\gamma, \beta)$ deviates little from unity. To visualize this, we have added two more horizontal planes, $f = 1 \pm 0.2$, above and below the pink plane $f = 1$, to represent the range where discrepancy between Hall effect and FET measurements is within ±20%. The result is shown as a top view of the function in Fig. 2b. The yellow and orange areas correspond to the parameter space, where Hall effect and FET measurements approximately match each other (within ±20%). This suggests that once a system is in the so-called "ideal" or "fully developed" Hall effect regime (that is, when $n_{Hall} \approx n_{FET}$ and $\mu_{FET} \approx \mu_{Hall}$, as in the case of Eq. 12b, or in the case of a pure band transport $\gamma = 1$), additional variations in the relative concentrations of hopping and band carriers or their mobilities will not easily lead to a significant change of the type of Hall effect (or a sizable discrepancy between Hall and FET measurements). For instance, when the fraction of band carriers in Fig. 2d is $\gamma > 0.8$, Hall effect will approximately match FET measurements irrespectively of particular mobility ratio $\beta$. This explains the empirical observation that if there is a good correspondence between Hall and FET results, it is rather difficult to disturb it by simply adding extrinsic disorder (physical defects).

To illustrate this experimentally, we have prepared a series of rubrene OFETs, based on single crystals subjected to various degrees of photooxidation (see Methods for details). It is well established that photooxidation creates charge traps in organic semiconductors, which is detrimental for carrier mobility in these materials[10,11]. We have performed *ac* Hall effect measurements in this series of devices, including a pristine (nearly trap-free) rubrene OFET, an OFET based on a lightly photooxidized rubrene crystal, and finally an OFET based on a severely photooxidized crystal (Fig. 3). All the devices had the same type of FET structure prepared using parylene-*N* as a gate dielectric (for details on this type of OFETs, see, e.g., ref.[27]). The Hall measurements reveal a seemingly surprising behavior: while, as expected, the drift carrier mobility in this series of OFETs decreases, following a sequence $\mu = 4$, 2.7 and $\sim 0.4$ cm$^2$V$^{-1}$s$^{-1}$, not only can we measure Hall effect, but we do observe a nearly proper (fully developed) Hall effect in all of these devices, including the one with the greatest amount of added disorder and the smallest mobility of 0.4 cm$^2$V$^{-1}$s$^{-1}$. Indeed, a good match between $\mu_{Hall}$ and $\mu_{FET}$, as well as $n_{Hall}$ and $n_{FET}$, is roughly maintained from pristine to severely photooxidized devices, even though $\mu$ is intentionally reduced well below 1 cm$^2$V$^{-1}$s$^{-1}$ (Fig. 3). Such "resilience" of Hall effect to the added disorder is understandable, given the shape of the function $f(\gamma, \beta)$ depicted in Fig. 2.

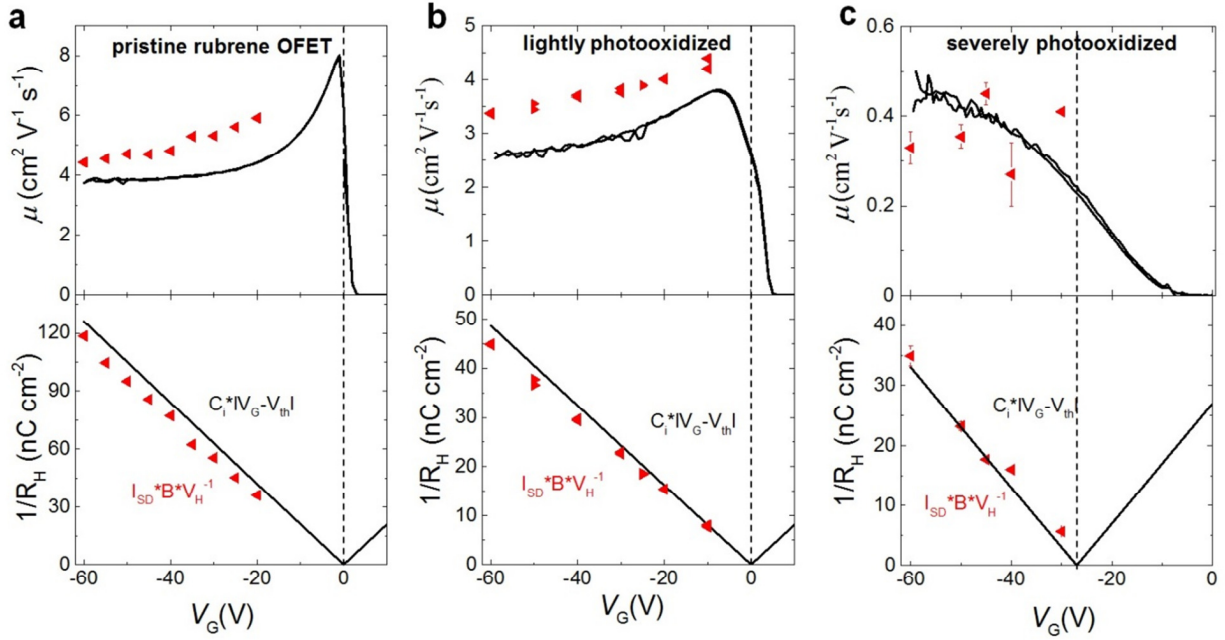

**Figure 3. Hall effect and FET measurements in a series of rubrene OFETs with intentionally varied degree of static disorder (density of defects).** Hall and FET data are shown with red triangles and black solid lines, respectively. The devices are fabricated on: (a) pristine, (b) lightly photooxidized, and (c) strongly photooxidized rubrene single crystals. The FET and Hall mobilities (the upper panels) and carrier densities (the lower panels) were determined in the linear regime of OFETs operation [7]. Threshold voltage $V_{th}$ is determined from the linear transfer characteristics recorded in the linear regime for each device ($V_{th}$ for pristine and lightly photooxidized devices were almost zero). A (nearly) fully developed Hall effect is observed in all the devices, even though the mobility in this series decreases from ~ 4 to ~ 0.4 cm$^2$V$^{-1}$s$^{-1}$.

It can be seen from the above equations that the Hall voltage compensation has some interesting consequences for the temperature dependence of carrier mobility, $\mu_{Hall}(T)$. In the case of a "strong hopping contribution" (Eq. 12a), when $n_{Hall} > n_{FET}$ and $\mu_{Hall} < \mu_{FET} < \mu_1$ (that is, when $\gamma \equiv n_1/n \ll 1$ is small, and $\beta < 1$ but not necessarily small), we obtain the following approximate expression for the Hall mobility from Eq. 8:

$$\mu_{Hall}(T) \approx \mu_1 \cdot \frac{\gamma}{\beta} = \mu_1 \frac{n_1}{n} \frac{\mu_1}{\mu_2} = \frac{\mu_1^2}{\mu_2} \frac{n_1}{n} \propto \frac{e^{\Delta_h/k_B T}}{T^{2\eta}} \cdot \frac{n_1}{n} \qquad (\eta, \Delta_h > 0). \qquad (13)$$

Here, the band mobility $\mu_1$ is expected to have a typical for band-like transport inverse temperature dependence $\mu_1 \propto T^{-\eta}$ with a power exponent $\eta \approx 1.5 - 2$, which has been experimentally observed and theoretically modeled in band-like organic semiconductors[28,29]. The hopping mobility $\mu_2$ is thermally activated, $\mu_2 \propto \exp(-\Delta_h/k_B T)$, where $\Delta_h > 0$ is a characteristic energy barrier for hopping in localized tail states, and $k_B$ is Boltzmann's constant. In the "intermediate hopping contribution" case (Eq. 12b), when $n_{Hall} \sim n_{FET}$, $\mu_{Hall} \sim \mu_{FET} < \mu_1$, and $\gamma = (\gamma - \gamma\beta + \beta)^2$ from Eq. 11, Eqs. 8 and 10 lead to the following approximate expression for the mobility:

$$\mu_{Hall}(T) \approx \mu_{FET}(T) \approx \mu_1 \cdot \gamma^{1/2} = \mu_1 \cdot \left(\frac{n_1}{n}\right)^{1/2} \propto \frac{1}{T^\eta} \cdot \left(\frac{n_1}{n}\right)^{1/2} \qquad (\eta > 0). \qquad (14)$$

First of all, these equations show that, contrary to the common expectation that temperature dependence of Hall mobility should be representative of the intrinsic band mobility, $\mu_1(T)$, this is not necessarily the case in a mixed transport regime: $\mu_{Hall}(T)$ above is the band mobility $\mu_1$ modified by the factors $\gamma/\beta$ or $\gamma^{1/2}$ in the cases described by Eqs. 13 and 14, respectively. Correspondingly, the measured temperature dependence of Hall mobility will differ from $\mu_1(T)$, because the factors are temperature dependent themselves. Second, the temperature dependences of the prefactors in front of $n_1/n$ and $(n_1/n)^{1/2}$ in Eqs. 13 and 14 are rather different. The prefactor in Eq. 13 has a strong "band-like" character that tends to make the temperature dependence of mobility band-like ($d\mu/dT < 0$). In Eq. 14, the prefactor also has a "band-like character", but weaker.

The fraction of band carriers $\gamma = n_1/n$ on the other hand must have a thermally activated temperature dependence. Indeed, in the model of carrier localization by dynamic disorder, we can think about individual carriers as *intermittently* existing in the two states: delocalized (band) or localized (hopping) states. The fraction of time an *average* carrier spends in one state or the other ($\tau_{band}$ and $\tau_{tail}$, respectively) depends on the strength of thermal fluctuations, thermal energy, position of the Fermi level with respect to the mobility edge, etc. Thus, a crystal should always have populations of band and hopping carriers, with their relative densities related to these time scales as: $n_{band}/n_{total} \equiv n_1/n = \tau_{band}/(\tau_{band} + \tau_{tail})$. Given the typical total carrier densities in OFETs ($n = 10^{11} - 10^{13}$ cm$^{-2}$) and the density of tail states in molecular crystals ($10^{14} - 10^{18}$ cm$^{-3}$eV$^{-1}$) [4,8,9], the Fermi level in the channel of an operating OFET typically occurs in the tail, separated from the mobility edge by an appreciable energy of a few hundred meV: $E_{ME} - E_F$ ~ 0.1 – 0.3 eV, which is much greater than k$_B T$ at room temperature. Thus, generation of band-like carriers will mainly rely on thermal excitation from the tail states to (and above) the mobility edge. Therefore, temperature dependence of $\gamma = n_1/n$ can be described by an activation exponent:

$$\gamma = n_1/n \approx e^{-(E_{ME}-E_F)/k_B T} \quad (15)$$

that competes with the "band-like" prefactors in Eqs. 13 and 14.

A counterintuitive prediction arising from Eqs. 13-15 is that such a competition shifts the character of $\mu_{Hall}(T)$ dependence toward a band-like type more significantly in the case of a strong hopping contribution (Eq. 13), compared to the intermediate hopping case (Eq. 14). Indeed, the extrema of functions in Eqs. 13 and 14 (with the parameters taken to be $\eta = 2$, $E_{ME} - E_F = 0.1$ eV and $\Delta_h = 0.026$ eV for certainty) occur at $T_{max}^{(13)} = (E_{ME} - E_F - \Delta_h)/4k_B$ ~ 210 K in the case of a strong hopping contribution, and $T_{max}^{(14)} = (E_{ME} - E_F)/4k_B$ ~ 290 K in the intermediate case. That is, $T_{max}^{(13)} < T_{max}^{(14)}$, suggesting that when hopping contribution is strong, a "band-like" $\mu(T)$ dependence persists to lower temperatures than that at moderate hopping. Given the Hall voltage compensation effect described above, such behavior makes sense, because the motion of hopping carriers quickly "freezes out" with cooling (and so does the effect of $V_{Hall}$ compensation), leading to a stronger apparent "band-like" tendency in $\mu_{Hall}(T)$.

Such a behavior has indeed been observed by J.-F. Chang *et al.* (see Fig. 1 c,d in [18]), where an apparently fully developed Hall effect, with $n_{Hall} = n_{FET}$ and $\mu_{Hall} = \mu_{FET}$, was found for 1,4,8,11-tetramethyl-6,13-triethylsilylethynyl pentacene (TMTES-P), suggesting that all the charge carriers in that system are band-like (and experience a classic Lorentz force), thus leading to a fully developed Hall effect. On the contrary, TIPS-P exhibited an underdeveloped Hall effect, with $n_{Hall} > n_{FET}$ and $\mu_{Hall} < \mu_{FET}$, suggesting that the charge carriers are only partially coherent. Yet, surprisingly, the mobility $\mu_{Hall}(T)$ in TMTES-P OFETs was purely thermally activated, which is inconsistent with the band-like character suggested by the ideal Hall effect, while TIPS-P exhibited $\mu_{Hall}(T)$ closer to a band-like at high temperatures, with a crossover to a thermally activated regime at about $T_{max}$ ~ 220 K. If $\mu_{Hall}(T)$ of TMTES-P exhibited a maximum, it would likely appear at a higher temperature (outside of the measured range). Our model shows that this apparent inconsistency is superficial, as it is resolved by considering the Hall voltage compensation effect described above.

We next study in more detail the effect of mixed transport regime on the temperature dependence of mobility $\mu(T)$ in the case of moderate hopping contribution (Eq. 12b). In this important case, an approximate match between Hall and FET carrier densities and mobilities is observed. We have carried out $\mu(T)$ measurements in OFETs based on intentionally photooxidized rubrene and pristine tetracene single crystals (Fig. 4). Since pristine rubrene OFETs have already been extensively studied, and a band-like (non-activated) $\mu(T)$ has been confirmed in these devices (see, e.g., [7] and refs. therein, and [16,28,29]), we will not be reproducing this case here. It's worth noting that recently developed high-resolution *ac* Hall measurements allowed resolving Hall effect in single-crystal tetracene OFETs[21]. These measurements demonstrated that Hall effect in this system is proper (fully developed), even though the carrier mobility is rather low, $\mu$ ~ 0.3 - 0.7 cm$^2$V$^{-1}$s$^{-1}$, similarly to the case of the photooxidized rubrene described above (Fig. 3b,c). Thus, both of these systems (photooxidized rubrene and pristine tetracene) seem to belong to the same category of systems with an intermediate hopping contribution (Eq. 12b), with $n_{Hall} \approx n_{FET}$ and $\mu_{Hall} \approx \mu_{FET} < \mu_1$. In this case, while the simplistic interpretation of Hall effect suggests that all the carriers are band-like, the experiment shows that the temperature dependence of mobility is, nevertheless, thermally activated (Fig. 4). This apparent inconsistency can be explained by the model considered above and specifically by Eqs. 14-15, according to which the mobility in this regime is a product of the two competing factors, a band-like one and a thermally activated one: $\mu(T) \propto T^{-\eta} \cdot \exp(-(E_{ME} - E_F)/2k_BT)$. Fitting $\mu(T)$ data in Fig. 4 with Eqs. 14-15, yields reasonable activation energy and power exponent, $E_{ME} - E_F$ ~ 0.3 eV and $\eta$ ~ 2.

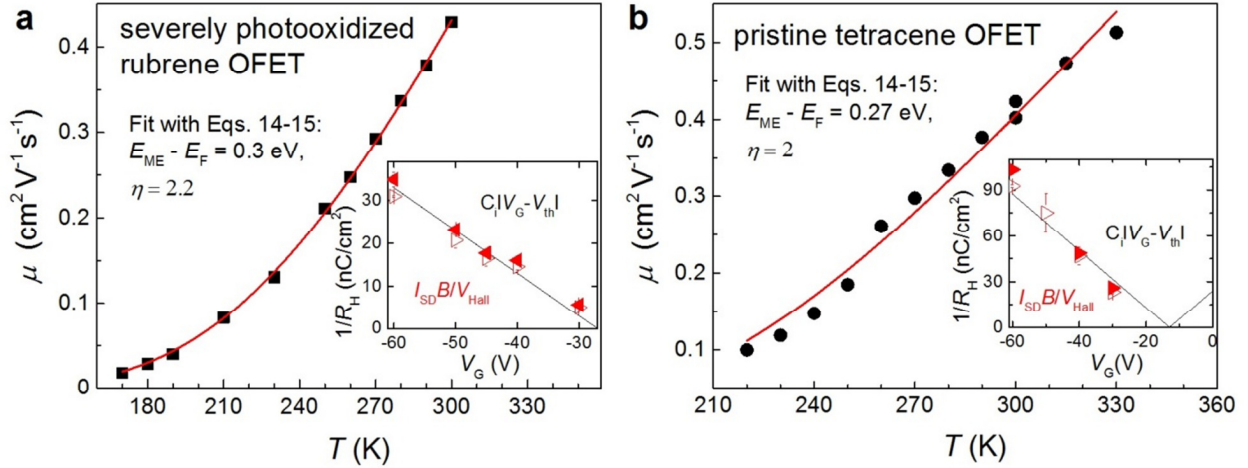

**Figure 4. Temperature dependence of carrier mobility in OFETs in a mixed (band and hopping) charge transport regime with moderate hopping contribution (Eqs. 12b, 14 and 15).** (a) OFETs based on severely photooxidized rubrene crystals, and (b) OFETs based on pristine, as-grown tetracene crystals. In these two cases, Hall effect is proper ($n_{Hall} \approx n_{FET}$ and $\mu_{Hall} \approx \mu_{FET}$), as illustrated in the insets by matching Hall (triangles) and FET (solid line) carrier densities, yet $\mu(T)$ is thermally activated (main panels). Such behavior can be rationalized within the model developed in this work: $\mu(T)$ is fitted with Eqs. 14-15 (solid red lines in the main panels), yielding reasonable activation energy, $E_{ME} - E_F \sim 0.3$ eV, and inverse-$T$ dependence power exponent, $\eta \sim 2$. All $T$-variable measurements have been carried our using a gated 4-probe technique [27].

It's worth noting that the carrier coherence factor, $\alpha \equiv n_{FET}/n_{Hall}$, introduced by T. Uemura *et al*. [17] can be expressed in our model in terms of parameters $\gamma$ and $\beta$ as:

$$\alpha \equiv \frac{n_{FET}}{n_{Hall}} = \frac{\mu_{Hall}}{\mu_{FET}} = \frac{\gamma}{(\gamma - \gamma\beta + \beta)^2}. \tag{16}$$

It's temperature dependence can be analytically or graphically investigated (Fig. 5). Plotting $\alpha$ as a 3D function of two variables, $\gamma$ and $\beta$, shows that when the initial value of $\alpha < 1$ (corresponding to a point on the turquoise surface below the pink plane $\alpha = 1$ in Fig. 5a), and temperature is reduced, thus causing the parameters $\beta = \mu_2/\mu_1 \propto T^\eta \cdot e^{-\Delta_h/k_B T}$ and $\gamma = n_1/n \propto e^{-(E_{ME} - E_F)/k_B T}$ to decrease (that is, move toward the origin), the point will typically move toward the spike (occurring near the origin), which corresponds to $\alpha$ increasing with cooling. A carrier coherence factor growing from 0.5 to ~ 0.65, when temperature is decreased from 300 to 155 K, has indeed been recently reported by T. Fukami *et al*. in Hall effect measurements of pentacene transistors[22].

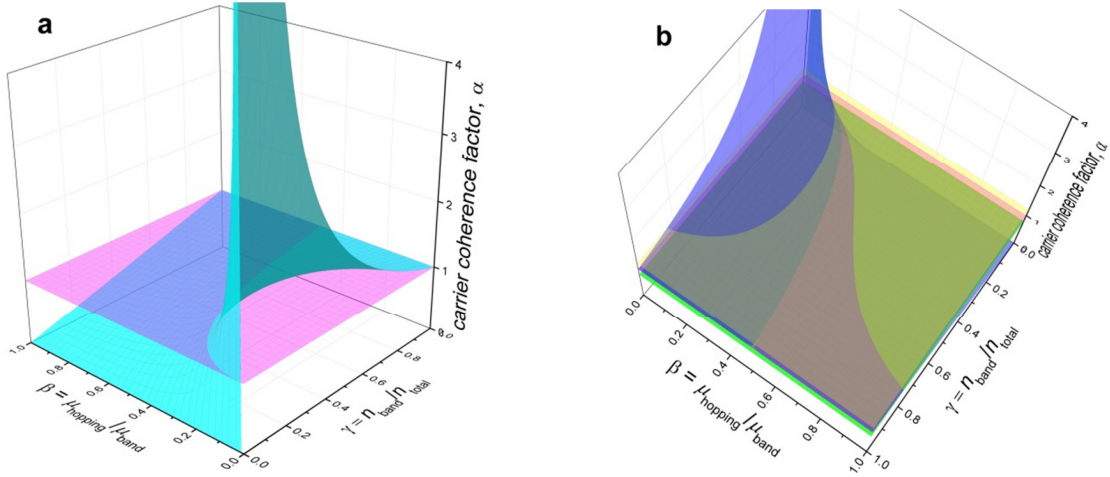

**Figure 5. Carrier coherence factor, α ≡ $n_{FET}/n_{Hall}$, given by Eq. 16.** (a) α plotted as a function of parameters γ = $n_1/n$ and β = $\mu_2/\mu_1$, representing the fraction of band-like carriers and the ratio of hopping and band mobilities, respectively (the turquoise surface). The constant α = 1 is given for comparison (pink plane). (b) The same function α with two added planes, α = 1 ± 0.2, viewed from above, showing the range of parameters, in which deviations from a fully developed Hall effect are within ±20% (the grey-pink areas). Origin file with the 3D function α can be downloaded from Supplementary Materials.

In conclusion, we show that Hall effect measurements in organic semiconductors in a mixed transport regime (with coexisting band and hopping carriers) may result in an incorrect estimate of the charge carrier mobility and density. The Hall mobility can be underestimated, compared to the true mobility of band-like carriers (as well as relative to the longitudinal FET mobility), while the Hall carrier density can be overestimated. In particular, the Hall carrier density can appear to be even greater than the total carrier density. This occurs because of a (partial) Hall voltage compensation due to the transverse drift of hopping carriers in the direction opposite to the Lorentz force. We have developed an analytical model of the Hall effect in organic semiconductors that expresses the Hall mobility $\mu_{Hall}$ and carrier density $n_{Hall}$, as well as the carrier coherence factor α in terms of the two microscopic parameters, the relative density of band carriers and the ratio of mobilities of hopping and band carriers. It has to be noted that in samples with substantial intrinsic or extrinsic disorder, we are likely to have a continuous mobility distribution, instead of just two "discrete" types of carriers, fast and slow, as we considered above. Thus, the simplified formulas derived here may not necessarily describe the actual disordered systems precisely. However, this model correctly captures the main tendencies observed in Hall effect measurements in such systems, including an underdeveloped Hall effect, peculiarities in the temperature dependence of carrier mobility and modifications of the carrier coherence factor α with temperature. This simple result emphasizes the caution that should be exercised while interpreting Hall effect data in disordered systems with coexisting band and hopping carriers.

**Methods**

    **Device fabrication.** Rubrene and tetracene single crystals were grown using a physical vapor transport method in a flow of ultra-high purity He gas, at sublimation temperatures of 320 and 280 °C, respectively. Electrical contacts of OFETs were prepared on (*a*,*b*) facets of the crystals with an aqueous solution of colloidal graphite (Ted Pella). Parylene-*N* was used as a gate insulator. In tetracene FETs, a very thin under-layer of Cytop (< 30 nm) was spin-coated on the crystals before depositing a much thicker layer of parylene-*N* ($C_i$ = 1.7 nF·cm$^{-2}$) (for details see ref. 21). For preparation of photooxidized rubrene OFETs, bare single crystals were photooxidized in a chamber filled with an atmosphere of ultra-high purity $O_2$ gas for 12 min (lightly photooxidized) and 2 hours (severely photooxidized) under a white-light illumination with a 85 mW·cm$^{-2}$ Xenon lamp placed 10 inches away from the samples. After the photooxidation and contact deposition, the crystals were coated with parylene-*N* ($C_i$ = 0.8 and 1 nF·cm$^{-2}$ for lightly and severely photooxidized crystals, respectively). 35 nm-thick Ag gate was thermally evaporated through a shadow mask on top of the parylene dielectric defining a channel with typical aspect ratio of *L*/*W* = 1 - 3, *D*/*W* = 0.3 - 0.7, where *L* is the channel length, *W* is the channel width, and *D* is the longitudinal distance between the 4-probe voltage probes.

    **Measurements details.** All measurements, except for *μ*(*T*) in Fig. 4, reported in this manuscript were performed at room temperature in air. Keithley Source-Meters K-2400 and Electrometers K-6512 were used for FET measurements. Gate voltage sweep rate was 1 V·s$^{-1}$. For temperature variable measurements, we used a closed-cycle cryostat (Advanced Research Systems). For *ac* Hall measurements, we used a low noise voltage preamplifier 1201 (DL instruments), a lock-in amplifier SR-830 (Stanford Research), a current source K-6221 (Keithley), and electrometers K-6514 (Keithley). *ac* magnetic field of a frequency in the range 0.5 - 3 Hz and strength of 0.23 T (rms) was generated by a rotating assembly of strong permanent neodymium magnets.

**Acknowledgements.**

We are indebted to the Institute for Advanced Materials and Devices for Nanotechnology (IAMDN) of Rutgers University for providing the necessary facilities and resources to support this project. We acknowledge the financial support of this work by the National Science Foundation grant DMR-1506609. We thank Prof. Michael Gershenson for helpful discussions and Dr. Hyun Ho Choi for assistance with device fabrication.

**Author Contributions.**

V.P. designed the experiments. H.T.Y. performed device fabrication and measurements. H.T.Y. and V.P. analyzed the data. V.P. and Y.G. created the model. V.P. wrote the paper. All authors discussed the results and contributed to editing the paper.

**Additional Information.**

**Supplementary information** accompanies this paper at http://www.nature.com/srep

**Competing financial interests:** The authors declare no competing financial interests.